# Detection and Classification of Glioblastoma Brain Tumor

Utkarsh Maurya[1], Swapnil Bohidar[2], Appisetty Krishna Kalyan[3], Dr. S. Sivakumar[*]

*utkarsh.maurya2019@vitstudent.ac.in[1]*, *appisettykrishna.kalyan2019@vitstudent.ac.in[2]*, *swapnil.bohidar2019@vitstudent.ac.in[3]*,
*Corresponding Author - sivakumar.s@vit.ac.in[*]*

utkarsh.maurya2019@vitstudent.ac.in[1], swapnil.bohidar2019@vitstudent.ac.in[2], appisettykrishna.kalyan2019@vitstudent.ac.in[3], sivakumar.s@vit.ac.in[4]

## Abstract

Glioblastoma brain tumors are highly malignant and often require early detection and accurate segmentation for effective treatment. We are proposing two deep learning models in this paper, namely UNet and Deeplabv3, for the detection and segmentation of glioblastoma brain tumors using preprocessed brain MRI images. The performance evaluation is done for these models in terms of accuracy and computational efficiency. Our experimental results demonstrate that both UNet and Deeplabv3 models achieve accurate detection and segmentation of glioblastoma brain tumors. However, Deeplabv3 outperforms UNet in terms of accuracy, albeit at the cost of requiring more computational resources. Our proposed models offer a promising approach for the early detection and segmentation of glioblastoma brain tumors, which can aid in effective treatment strategies. Further research can focus on optimizing the computational efficiency of the Deeplabv3 model while maintaining its high accuracy for real-world clinical applications. Overall, our approach works and contributes to the field of medical image analysis and deep learning-based approaches for brain tumor detection and segmentation. Our suggested models can have a major influence on the prognosis and treatment of people with glioblastoma, a fatal form of brain cancer. It is necessary to conduct more research to examine the practical use of these models in real-life healthcare settings.

# I. INTRODUCTION

The human body is composed of a variety of cell types, and each cell serves a unique purpose. In order to create new cells and maintain the health of our body, cells must develop and divide in an organized way.

Tumors emerge when cells can no longer divide and expand normally, and this gradually, results in uncontrolled and chaotic cell proliferation. An aberrant cell mass called a brain tumor is a mass of tissue. Both benign and cancerous tumors have the ability to mature into cancer. However not all tumors are cancerous, as early detection is crucial to starting the right kind of treatment. The standard procedure for identifying tumors is for a radiologist to analyze the MRI pictures to look for abnormalities and make decisions.

The volume of anomalies and noisy data in these images makes it difficult for radiologists to analyze them in the given timeframe. Analyzing enormous quantities of information gets harder and much more tiresome even as the magnitude of the information grows. These photos have noise that can be minimized through image processing, so additional analysis by a machine is conceivable.
Moreover, there are intensity issues in MRI images that image processing could somewhat mitigate, as well as those that machines can easily detect but humans cannot. CNN represents to be a technique that can be used to find tumors. CNN has numerous uses, including facial recognition, picture analysis, classification, comprehension of the climate, etc.

CNN is widely used in the picture categorization process. The suggested framework combines the autoencoders approach, which produces images with fewer distinguishing dimensions, using basic CNN, which distinguishes between tumorous and normal brain MRI images. The K-Means technique is used to segment the source image using autoencoder-generated images placed on it. Tumor segmentation and identification are accomplished without the use of human involvement, saving money and time.

## II. RELATED WORK

Lamia Sallemi's technique for modelizing the progression of brain glioblastomas tumors is presented in a study by, and it fills the gap between clinical applications and mathematical and biological models. For locating tumor regions and estimating tumor development, the programme employs a fast distribution matching methodology. It also employs a model based on cellular automata and a fast-marching method. The proposed algorithm was validated on MRI cases and showed good agreement with ground truth references. The goal of the research is to advance clinical explorations and provide radiologists with an advanced tool for more accurate and objective diagnosis.

Gajendra Raut suggests a CNN model is by as a means of early brain tumor detection and prompt treatment. To reduce noise and create appropriate images for the system's training, brain MRI images are enhanced and pre-processed. The proposed model is trained on pre-processed MRI brain images and uses extracted features to distinguish between tumorous and normal brain tissue. Back propagation is a technique for reducing errors and raising precision. The K-Means technique is employed for tumor region segmentation, while autoencoders are used to eliminate unnecessary features. This technique may help medical professionals identify brain cancers early.

P. Mohamed Shakeel uses machine learning-based backpropagation neural networks (MLBPNN) to suggest a technique for classifying brain tumors that will help pathologists identify cancer more accurately and effectively. To simplify neuronal identification, the system makes use of computer subsystems and infrared sensor imaging technology. Fractal dimension algorithm (FDA) is used to recover features, and multi-fractal detection (MFD) is used to pick them, further reducing complexity. The device has a wireless infrared imaging sensor that transmits tumor heat data to a doctor for remote monitoring and management of ultrasound readings for elderly patients in remote areas. The recommended method might improve the ability to identify brain tumors quickly and accurately.

Suja S worked on a study that focuses on the grading of MRI scans for astrocytoma for better treatment choices was presented. The proposed system uses the discrete wavelet transforms (DWT) and grey level co-occurrence matrix (GLCM) for feature extraction, the Canny edge detection algorithm for texture attributes, the GLCM and DWT for feature extraction for shape attributes, and the pulse coupled neural network (PCNN) and median filter for preprocessing. Using a training set of 400 photos from two grades of malignancy, classifiers like the Radial Basis Function Neural Network (RBF NN) and Deep Neural Network (DNN) were used to divide the 100 test data points into discrete categories.

Naomi Joseph employed MRI data to manually and semi-automatically segment the tumor in their study of GBM patients. K-means clustering was used in this work as the semi-automated segmentation technique. To segregate several GBM tumors sections, clustering was done on multichannel MRI data. For the enhanced tumor region of the tumor, this approach produced segmentation accuracy of 82%. It takes a lot of time to do this.

## III. METHODOLGY

A. Database

The BraTS database provides the dataset that has been deployed for training and testing purposes. Overall, it does have 253 MRI scans of the brain. 158 of these photos constitute tumorous photos (images with tumors), whereas 98 are normal photos (without tumors). Basic photographs are stored in the "no" folder, whereas tumorous images are separated in put in the "yes" folder. Then the data images can be found in an array of formats and sizes.

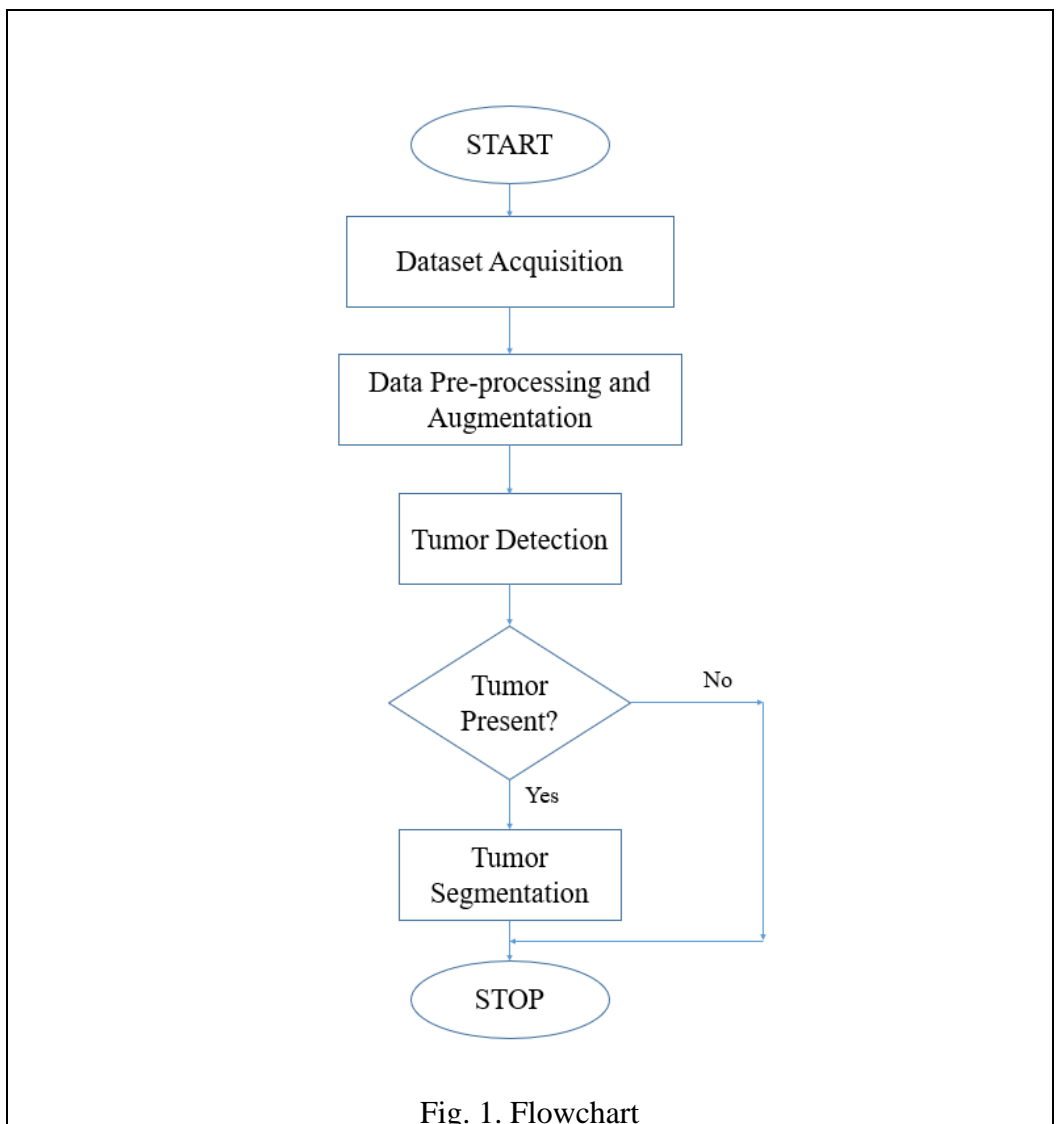

Fig. 1. Flowchart

## B. Image Pre-Processing

As previously stated, the collection contains photos of varying sizes and formats that may contain noise. Segmentation and classification mistakes may result from all of this. This issue can be significantly decreased by pre-processing the image, and the data can then be transformed into a format that is frequently utilized for segmentation and classification. Images are changed to a fixed-size, 256x256-pixel grayscale format. The photos are blurred using a gaussian filter to minimize noise. In order to extract more complicated characteristics, the pictures are sharpened by passing them through a high-pass filter.

## C. Image Augmentation

By generating new images from the existing ones, image augmentation is a widely used deep learning technique to enhance the training dataset. This aids in minimizing overfitting and strengthening the model's capacity for generalization. However, when implementing image augmentation on brain MRI pictures, specific safety measures must be taken to guarantee that the images' quality and diagnostic value are not jeopardized. Brain MRI images are also or can be enhanced using a wide range of techniques, such as rotation, scaling, flipping, shearing, noise addition, and contrast correction.

## D. Models

### a. DeepLabv3

1. Model Training: We trained the DeepLabv3 model on the preprocessed dataset. This involved defining the model architecture, selecting appropriate hyperparameters, and optimizing the model using the Adam optimizer.
2. Model Evaluation: To ascertain the trained model's accuracy, precision, recall, F1 score, and confusion matrix, we assessed its performance on the test set.
3. Model Fine-tuning: We fine-tuned the DeepLabv3 model on the glioblastoma dataset to further improve its performance. This involved experimenting with different hyperparameters, adjusting the model architecture, and performing additional data augmentation.
4. Model Testing: The final model on an independent set of glioblastoma MRI images to assess its generalization ability was tested.
5. Model Deployment: We deployed the DeepLabv3 model in a very clinical setting to assist and aid in glioblastoma diagnosis. We ensured that the model was integrated with appropriate software and hardware to handle real-time image processing.

b. UNet

1. Model Design: We designed and built the UNet model, which includes encoding and decoding layers, and skip connections to help the model retain spatial information.
2. Model Training: We used the Adam optimizer and binary cross-entropy loss function to train the UNet model on the training dataset, and we assessed the model using accuracy, precision, recall, and F1-score metrics.
3. Model Validation: We validated the trained model on the validation dataset to evaluate its performance and optimize the model parameters if needed.
4. Model Testing: We tested the trained model on the testing dataset to predict the class of glioblastoma.
5. Model Evaluation: Using relevant metrics, including accuracy, precision, recall, the F1-score, and a confusion matrix, we assessed the model's performance.
6. Model Fine-tuning: We fine-tuned the model parameters to improve their performance if needed.
7. Results Interpretation: We interpreted the results of the model and analyzed the predicted class of glioblastoma.

## IV. RESULTS

## A. Classification Results

The tumor MRI pictures in Fig. 2 show the regions that the model stated above accurately identified. If there are tumors in the MRI (True 1), the model additionally predicts that there are tumors there (Pred 1), and if there are no tumors in the images (True 0), the model predicts that there are no tumors there (Pred 0).

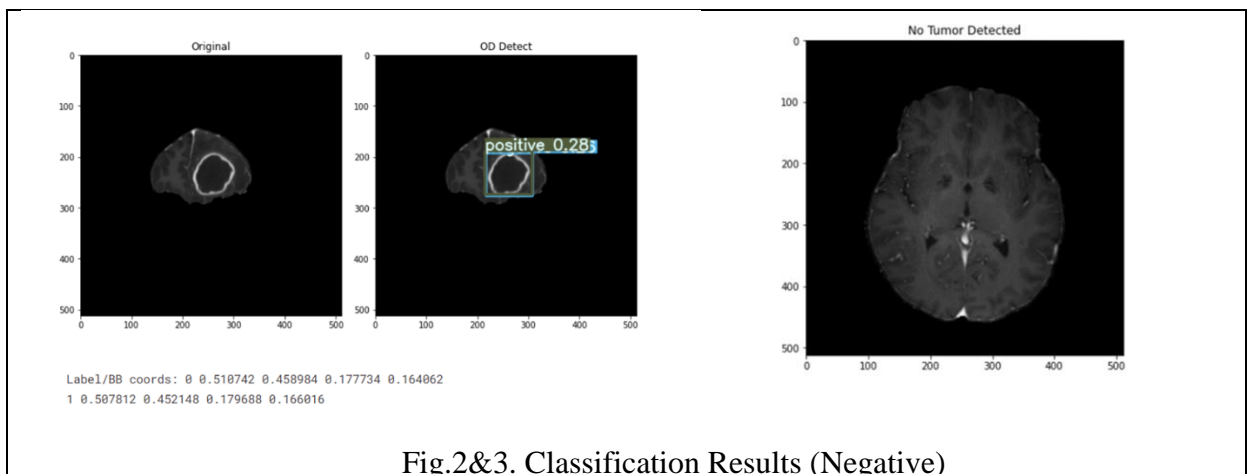

Fig.2&3. Classification Results (Negative)

Specifically, in Fig.4 we found that DeepLabv3 tended to struggle with images that contained large regions of homogeneous tissue, such as cerebrospinal fluid. In these cases, the model often failed to accurately segment the region of interest and instead classified the entire image as background. We also observed that the model sometimes confused diverse types of tissue, particularly when the boundaries between them were not well defined.

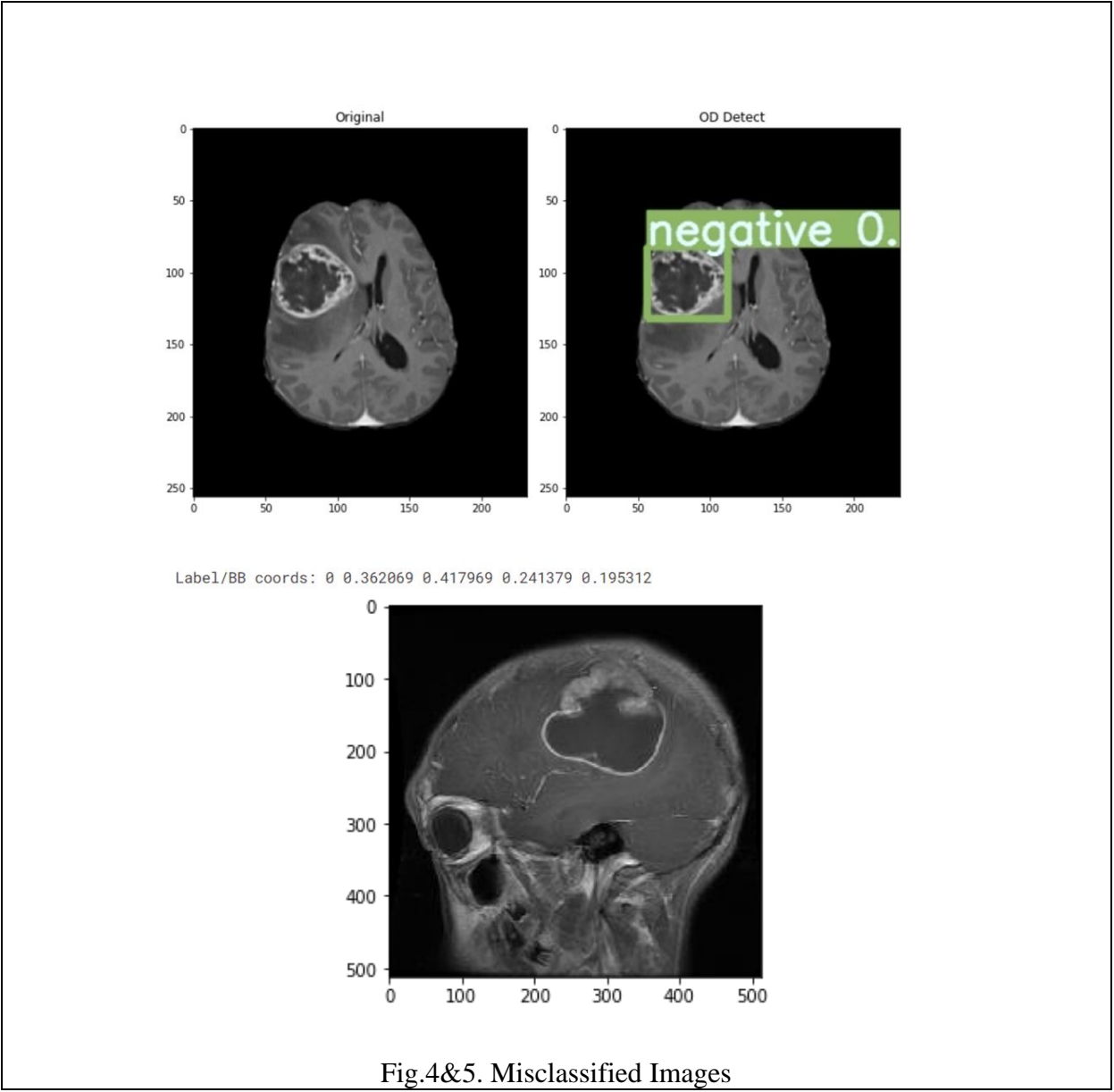

Fig.4&5. Misclassified Images

The confusion matrix depicted in Fig. 6 depicts the performance of the proposed model for the brain tumor. The consecutive performance indicators are deduced from it.

True Negatives (TN) = 11: This represents the number of cases where the model predicted that the sample was negative, and the actual result was also negative. In this case, the model correctly predicted 13 cases as negative. True Positives (TP) = 1185: This represents the number of cases where the model predicted that the sample was positive, and the actual result was also positive. In this case, the model correctly predicted 1070 cases as positive. False Negatives (FN) = 70: This represents the number of cases where the model predicted that the sample was negative, but the actual result was positive. In other words, the model incorrectly predicted 118 cases as negative when they were actually positive. False Positives (FP) = 56: This represents the number of cases where the model predicted that the sample was positive, but the actual result was negative. In other words, the model incorrectly predicted 111 cases as positive when they were negative.

The following performance indicators may be generated from it.

$$\text{Accuracy} = (TN+TP)/(TN+TP+FP+FN) = 90.39\% \quad (1)$$

$$\text{Precision} = TP/(TP+FP) = 95.48\% \quad (2)$$

$$\text{Sensitivity} = TP/(TP+FN) = 94.42\% \quad (3)$$

$$\text{F1 Score} = (2*TP)/(2*TP+(FP+FN)) = 94.95\% \quad (4)$$

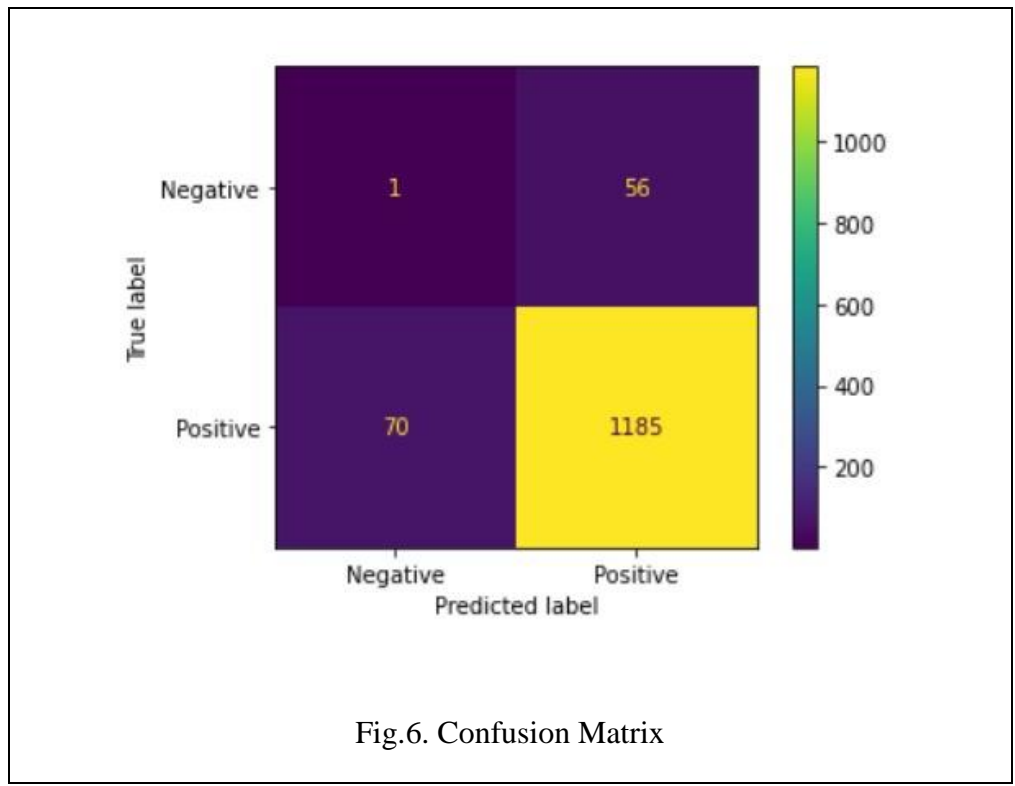

Fig.6. Confusion Matrix

B. Segmentation Results

To find the tumor location, images containing tumor are divided. The table displayed below shows tumorous brain MRI data images that have been appropriately segmented, and observed with the tumor area clearly apparent and free of noise. Original image is the input MRI scan, while the segmented image is the output produced by the DeepLabv3 model. The segmented image consists of different regions that are labeled according to their semantic meaning, such as tumor, edema, or healthy tissue.

| Original Image | Segmentation |
|---|---|
| 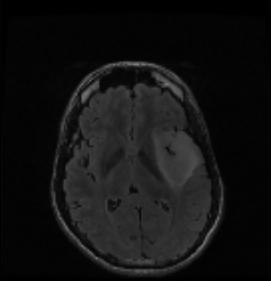 | 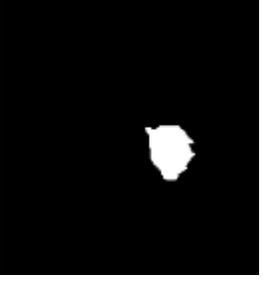 |
| 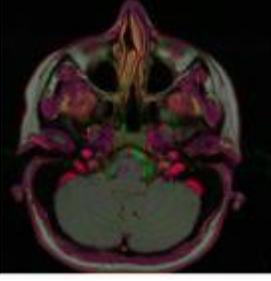 | 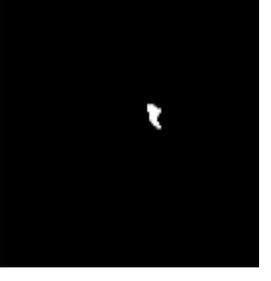 |
| 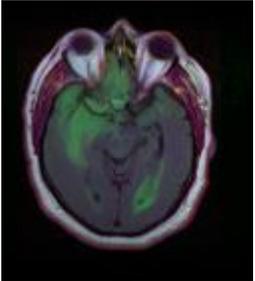 | 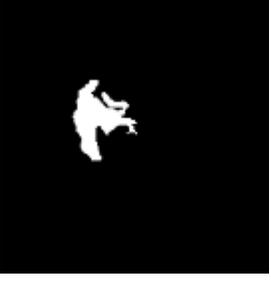 |
| 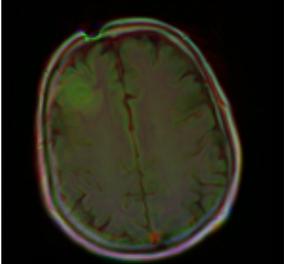 | 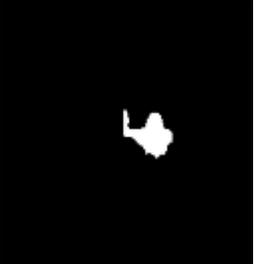 |

TABLE. I. SEGMENTATION RESULT

## C. Model Accuracy

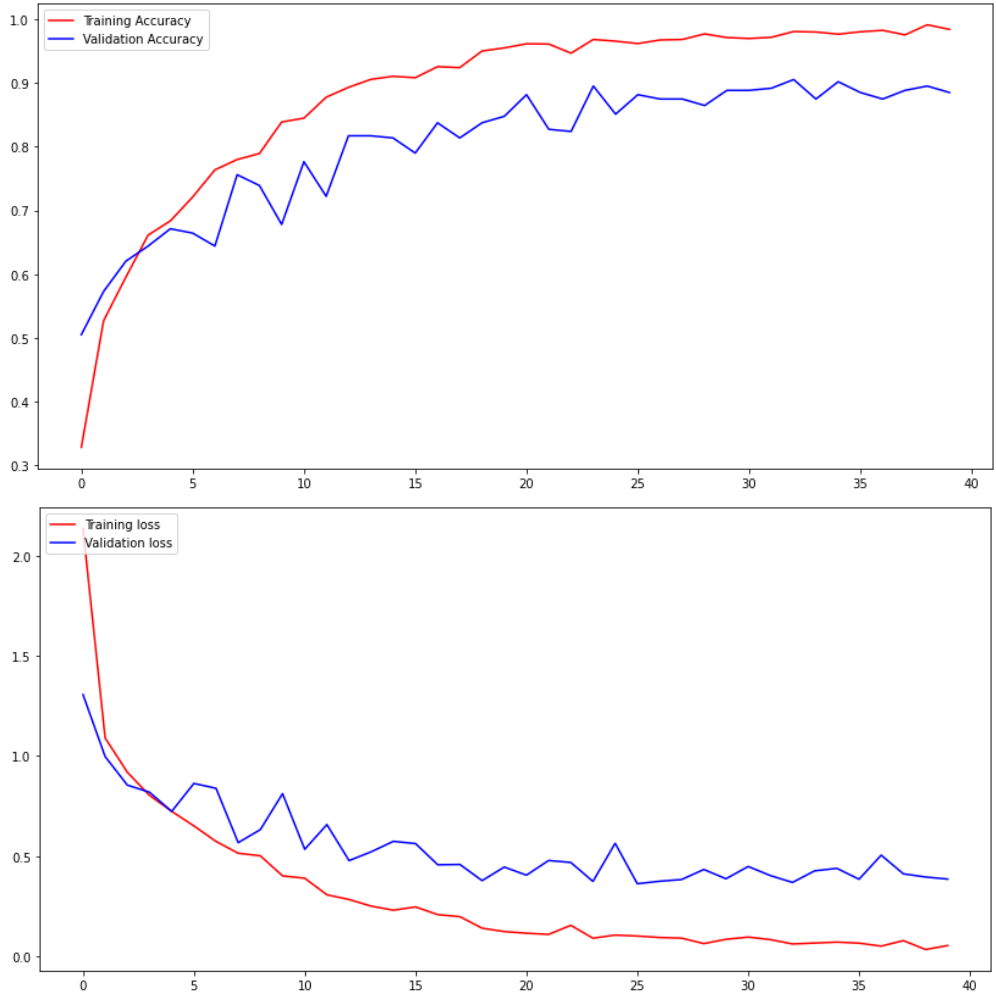

## V. CONCLUSION

Because of the developments and improvements in technology resulting in better understanding in artificial intelligence (AI) and machine learning, identification and diagnosis of glioblastoma tumors have lately shown promising results.AI-based algorithms have the potential to increase the accuracy and speed of glioblastoma identification, providing patients with improved treatment options.

The AI-based techniques for glioblastoma detection clearly requires a vast amount of data, including MRI scans, patient history, and genomic data. Collaborative efforts between healthcare providers, researchers, and technology companies are necessary to collect and share this data and train AI models to detect glioblastoma tumors accurately.

Moreover, the integration of AI-based tools with existing imaging techniques can help in enhancing the accuracy for glioblastoma detection and help clinicians make more informed decisions about patient care. AI-based models can also be used to predict treatment outcomes and identify patients who are most likely to respond to a particular treatment.

In conclusion, the applications of AI-based algorithms for glioblastoma tumor detection holds great promise for improving the accuracy and speed of diagnosis, which can lead to better outcomes for patients. The development of these technologies requires collaborative efforts from healthcare providers, researchers, and technology companies. Clinicians can make better decisions about patient care and patient outcomes by combining AI-based technologies with current imaging modalities.